# CLIC-LHC Based FEL-Nucleus Collider: Feasibility and Physics Search Potential

H. Braun[1], R. Corsini[1], J. P. Delahaye[1], E. Guliyev[2], A. Ozcan[3],
S. Sultansoy[2,3], O. Yavas[4], S. Yigit[4,5]

[1]CERN CH-1211, Genève 23, Switzerland
[2]Inst. of Phys. Acad. of Sciences, H. Cavid Ave. 33, Baku, Azerbaijan
[3]Gazi University, Fac. of Arts and Sci. Dept. of Physics
06500, Teknikokullar, Ankara, Turkey
[4]Ankara University, Fac. of Eng. Dept. of Eng. of Physics
06100, Tandogan, Ankara, Turkey
[5]Turkish Atomic Energy Authority, 06530, Lodullu, Ankara, Turkey

**Abstract**

The feasibility of a CLIC-LHC based FEL-nucleus collider is investigated. It is shown that the proposed scheme satisfies all requirements of an ideal photon source for the Nuclear Resonance Fluorescence method. The tunability, monochromaticity and high polarization of the FEL beam together with high statistics and huge energy of LHC nucleus beams will give an unique opportunity to determine different characteristics of excited nuclear levels. The physics potential of the proposed collider is illustrated for a beam of Pb nuclei.

**1 Introduction**

The investigation of collective excitations, especially electric and magnetic dipole vibrations of nuclei plays an important role in determination of nuclear structure. The Nuclear Resonant Fluorescence (NRF) is the most efficient method to study these excitations (see [1] and references therein). However, the capacity of traditional NRF experiments for study of higher line excitations are limited by several factors [2].

Recently, a new method, namely FEL-Nucleus Collider, has been proposed [3,4] which satisfy all requirements on ideal photon sources [1]: high spectral intensity, good monochromaticity, tunability in a broad energy range and high degree of linear polarization. For these reasons, FEL-nucleus colliders will be powerful tools for experimental determination of energy levels, decay widths, spin and parity of excited nuclei. The physics search potential of FEL-nucleus collider was demonstrated for the $^{154}$Sm nucleus excitations in [5], where TTF FEL is considered as the source of photon beam. The advantages of FEL-RHIC collider with respect to traditional NRF methods are demonstrated in [6] for some excitation levels of $^{232}$Th nucleus.

In this paper, we investigated the feasibility and physics search potential of CLIC-LHC based FEL-Nucleus collider. It is shown that the pulses of CLIC drive beam can be used to obtain FEL beam and LHC nucleus beam energies well suited to cover the typical excitation energies for different nuclei. In section 2, the disadvantages and limitations of traditional NRF methods are discussed. The requirements of an ideal photon source for nuclear spectroscopy are presented and a FEL is compared with traditional γ sources. The main parameters of a FEL-nucleus collider are discussed. In section 3, the time structure of CLIC drive beam, main characteristics of CLIC based FEL beam and matching of the beam structures for CLIC and LHC are investigated. In section 4, the physics search potential of the CLIC-LHC based FEL-nucleus collider is demonstrated considering excitation levels of Pb nucleus.

**2 Nuclear Spectroscopy, Traditional Photon Sources and FEL-Nucleus Colliders**

The data on electric and magnetic dipole excitations have revealed exciting new insights into nuclear structure. The study of these excitations gives valuable information about nuclear structure and nucleon-nucleon forces at low energy. For example, low-lying dipole excitations (so-called M1 scissors mode) and dipole excitations near particle threshold in transition and deformed even-even nuclei are of considerable interest in modern nuclear structure physics. Both M1 and E1 excitations were studied systematically to test nuclear models. Unfortunately, spin and parity of a lot of excitations are not determined with existing experimental facilities. Therefore, sensitive experimental tools are needed to investigate electromagnetic dipole and other higher order excitations in addition to well known Nuclear Resonance Fluorescence methods [1].

Recently electro- magnetic dipole response E1 and M1 have been studied using real photons as projectiles. The photon scattering method gives complete and model independent information about dipole strength distributions. However, the capacity of

NRF experiments for study of higher- lying excitations is limited by many factors. For example:

- Measuring the polarization of scattered photons using Compton polarimeters has been very successful at energies near and below 3 MeV. However, this technique becomes difficult at excitation energies exceeding 4 MeV because the analyzing power of the Compton scattering process is energy dependent and approaches zero at these energies.

- The rather weak interaction between the photons and the target material. Therefore, for typical obtainable photon currents of about $10^6$ ?/(s keV) at the target position one needs large amounts of isotope enriched target material (500mg typically)

There are several methods to produce photons for low energy photon scattering experiments. An ideal photon source for such experiments should have the following characteristics [1]:

- High spectral intensity $I = N_\gamma / eVs$ (number of photons per energy bin and second).
- Good monochromaticity ($\Delta E_\gamma / E_\gamma$).
- Tunable in a broad energy range.
- High degree of linear polarization ($P_\gamma \approx 100\%$).

Unfortunately up to now there are no such ideal sources available fulfilling all these requirements. Therefore, diverse photon sources have been applied in low energy photon scattering depending on the special experimental requirements and aims intended in the investigations. Therefore, only with the advent of a new experimental facility with improved characteristics it will be possible to investigate in detail the fine structure of the magnetic and electric response. FEL-nucleus colliders will have all the required characteristics due to the use of a FEL beam as photon source.

## 2.1 Free Electron Lasers

Free electron lasers can produce a photon beam in a wide region of the electromagnetic spectrum from infra-red to hard X-ray using relativistic electron beams with small emittance and undulator magnets [7-9]. In the SASE FEL, a high quality beam of relativistic electrons is injected into an undulator, i.e. a straight set-up of magnetic dipoles with equal field strength and alternated polarities [8]. These magnets are placed with a distance $\lambda_u$ between two identical poles. The beam is deflected by each of the magnetic fields and travels through the undulator on a sinusoidal trajectory. Since the electrons have a high velocity, they emit radiation in a small cone, in the forward direction. The strength of the interaction between the electron beam and the radiation field is characterized by the FEL parameter $r$

$$r \propto \gamma^{-1} J^{1/3} B_u^{2/3} \lambda_u^{4/3} \qquad (1)$$

where $\gamma$ is the electron energy in units of $m_e c^2$, J is the electron peak current density, $B_u$ is the rms field: if $B_0$ is the peak on-axis field, then $B_u = B_0/\sqrt{2}$ for a planar undulator and $B_u = B_0$ for a helical undulator and $\lambda_u$ is the undulator pole wavelength. Many important properties of the FEL directly depend on $r$. For example, if $z$ is the distance along the undulator, then the growth of the radiation power before saturation is given by:

$$P(z) \approx P_0 e^{z/L_g} \qquad (2)$$

where the gain length $L_g$ can be expressed as:

$$L_g \approx \frac{\lambda_u}{4\pi\sqrt{3}r} \qquad (3)$$

The parameter $r$ also gives the power extraction efficiency at saturation:

$$P_{sat} \approx r P_{beam} \qquad (4)$$

The wavelengths to be emitted depend primarily on the energy of the beam, as well as the period and the strength of the magnetic field in the undulator:

$$\lambda_r \approx \frac{\lambda_u}{2\gamma^2}(1+a_u^2) \qquad (5)$$

where the undulator parameter $a_u$ is defined as:

$$a_u = \frac{e\lambda_u B_u}{2\pi m_e c} \approx 0.934 \lambda_u(cm) B_u(T) \qquad (6)$$

This resonance condition means that the difference in path length between the electrons and the radiation field over one undulator period $\lambda_u$ must be equal to the wavelength $\lambda_r$ of the spontaneously emitted light. It also means that the wavelength is tunable by changing the undulator parameters, $\lambda_u$ and $B_u$, or the energy.

The equation (5) is only valid for electrons which have energy $\gamma mc^2$ and which travel on the central trajectory in the undulator. However, in reality, there is an energy spread within the electron bunch, which may broaden the line width $\Delta f_r = f_r$ (i.e. the frequency spread of the radiation), and even interfere with the power gain mechanism of the FEL. In order to obtain the maximum possible gain, one should have:

$$\Delta\gamma/\gamma \leq \rho \qquad (7)$$

The transverse betatron motion of the electrons may also be responsible for the broadening of the line width, because a spread in transverse velocities leads to a spread in the longitudinal velocities as well, which induces an additional effective energy spread. Therefore, the beam emittance is also a critical parameter, especially for short radiation wavelengths.

## 2.2 FEL-Nucleus Collider

The accelerated fully ionized nuclei will "see" the few keV energy FEL photons as a laser beam with MeV energy [3, 4]. A schematic view of FEL-Nucleus collider is given in Figure 1.

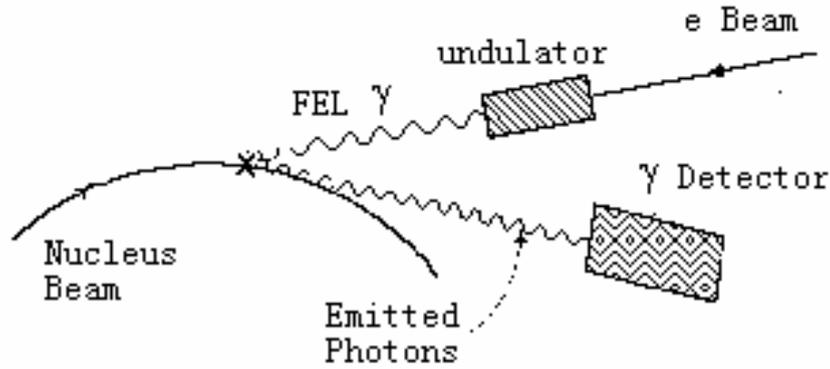

Figure 1. The schematic view of FEL-Nucleus collider.

Due to good monochromaticity ($\Delta E_\gamma/E_\gamma \sim ? <10^{-3}$-$10^{-4}$) with a typical obtainable number of photons of the order of $10^{13}\gamma$/bunch, tunability and better polarization ($P_\gamma \sim 100\%$) of the FEL beam. FEL-nucleus colliders can be successfully used to investigate nuclear excitations with low multi polarity in a wide energy region. The needed energy of FEL photons can be expressed as:

$$w_0 = \frac{E^*}{2\gamma_A} = \frac{A}{Z}\frac{E^*}{2\gamma_p} \qquad (8)$$

where $E^*$ is the energy of the corresponding excited level, A and Z are the atomic and mass numbers of the nucleus, $\gamma_A$ and $\gamma_p$ are the Lorentz factors of the nucleus and proton, respectively.

The luminosity of the FEL-nucleus collider can be expressed as,

$$L = \frac{n_g n_A}{4\pi \sigma_x \sigma_y} n_b f_{rep} \qquad (9)$$

where $n_\gamma$ and $n_A$ are the number of particles in FEL and nucleus bunches, respectively, $\sigma_x$ and $\sigma_y$ are the transverse beam sizes, $n_b$ is the number of bunches per FEL pulse, $f_{rep}$ is pulse frequency.

The cross section for the resonant photon scattering is given by the Breit-Wigner formula:

$$\sigma_{res}(\gamma,\gamma') = \frac{\pi}{E^2} \frac{2J_{exc}+1}{2(2J_0+1)} \frac{\Gamma^2}{(E-E_R)^2 + \Gamma^2/4} \qquad (10)$$

where E is the c.m. energy of the incoming photon (in our case it is very close to that in the rest frame of the nucleus), $J_{exc}$ and $J_0$ are spins of the excited and ground states of the nucleus and $\Gamma$ is the total width of the excited states.

Since the width of the excited states are smaller than the energy spread of colliding beams, the approximate value of the averaged cross sections has been found to be

$$\sigma_{ave} \approx \sigma_{res} \frac{\Gamma}{\Delta E_g} \qquad (11)$$

where $\Delta E_\gamma$ is the energy spread of the FEL beam in the nucleus rest frame; $\Delta E_\gamma/E_\gamma \leq 10^{-4}$, $E_\gamma \approx E_{exc}$. The number of the produced excited levels is given by

$$R = L\sigma_{ave} \qquad (12)$$

A comparison of traditionally used photon sources [1] with FEL is presented in Table 1. The spectral intensity of FEL beam is given for CLIC drive beam FEL (see following section). The target mass is determined by the product of nucleus bunch population and number of bunches in ring.

Table 1. Main characteristics of different photon sources from ref. [1] in comparison with FEL (last row).

| Photon Source | Spectral Intensity [γ/s·eV] | ΔEγ/Eγ [%] | Pγ [%] | Target Mass M [g] |
|---|---|---|---|---|
| Compton Backscattered (CB) | 0.15 | 2.7 | 100 | 70 |
| Bremsstrahlung (Polarized) | 20 | Cont. | 10-30 | 5 |
| Bremsstrahlung (Unpolarized) + CB | 1000 | Cont. | 10-20 | 5 |
| Bremsstrahlung (Unpolarized) | 1000 | Cont. | 0 | 1-2 |
| Free Electron Laser | >$10^{16}$ | 0.01 | 100 | $10^{-10}$ |

**3 CLIC Drive Beam Based FEL-LHC Collider**

It is well known that energy range of electromagnetic low multi-polarity excitations is 2-20 MeV. The corresponding FEL energies can be obtained from Eq. (8). For example, we obtain a wavelength range from 0.4 nm to 4 nm (0.34-3.4 keV photon energy) for Pb excitations, taking into account that $g_{Pb} \approx 2941$ for LHC [12, 13]. According to Eqs (5) and (6), to obtain a FEL with these energies we need $E_e \approx 1.2 - 4\,\text{GeV}$ for $l_u = 2.5\,\text{cm}$ and $B_u = 0.43\,\text{T}$.

In [3,4], a FEL driven by a superconducting linac has been considered. In the case of LHC, and if CLIC is eventually built at CERN [14], it would be interesting to explore the possibility of using CLIC components to accelerate the FEL electron beam. It must be noted that such an FEL would also be of interest for applications other that the nucleus collider.

Potential advantages of a CLIC-based FEL would be the (partial) use of existing equipment, and the possibility of reaching a high acceleration gradient, resulting in a shorter linac. On the other hand, as pointed out in previous work [10], the use of high frequency RF for acceleration is disadvantageous in terms of beam energy spread, which is required to be small by FEL physics. A superconducting linac is also in general better in terms of average brightness, and the time structure of the beam can be more easily matched with the time structure of the LHC ion beam.

In a CLIC-based FEL, one option would be to use additional pulses from the main beam, at low energy, in order to drive the FEL. This option is not particularly attractive, since the main beam before the damping rings has a relatively high emittance and, while after the damping ring the emittance is very small (but asymmetric in both planes), the electron bunches are long and therefore not adapted to FEL operation. Another problem comes from the fact that the main beam energy in the damping ring is about 2 GeV, while the FEL would need energies ranging from 1 to 4 GeV. Furthermore, it will be complicated to change the beam energy if parallel operation with CLIC is required.

A more promising option, which we will explore in the following, is to use a fraction of the CLIC drive beam to power a high-frequency, normal-conducting linac dedicated to FEL operation. A natural choice of the linac frequency would be 15 GHz, which corresponds to the drive beam bunch repetition frequency (in CLIC the power is produced at the 2$^\text{nd}$ harmonic of the drive beam, 30 GHz). The linac will have a lower gradient and/or will be less efficient than the CLIC main linac, but the energy spread can be made smaller, thus satisfying FEL requirements (typically $s_p < 10^{-3}$). Following a previous investigation [10], the minimum energy spread achievable for 30 GHz acceleration of a 1 nC, 50 µm rms long electron bunch is of the order of $10^{-3}$, therefore 15 GHz acceleration should give some useful margin.

**3.1 Luminosity and Time Structure Matching**

The LHC ion beam is composed of bunches with rms duration of 250 ps, separated by 100 ns. In a CLIC-based FEL, while the repetition rate is of the order of 200 Hz, the bunch spacing is linked to the time structure of the RF pulses that can be produced by the drive beam. Using only a drive beam pulse per cycle (~ 2 % of the total CLIC power) to accelerate a single electron bunch ($f_{rep}$ = 150 Hz, $n_b$ = 1) and assuming $s_x = s_y = 30$ µm, $n_A = 7\ 10^7$ and $n_g = 2.5\ 10^{13}$ we can obtain a "minimum" value for the luminosity: $L = 2.3\ 10^{27}$ cm$^{-2}$s$^{-1}$.

Although the luminosity thus obtained is not completely negligible, the situation is far from optimum. Some improvements can be expected by a somewhat better matching of the nucleus beam and the gamma beam sizes (16 µm and 30 µm, respectively); but the highest gain can be obtained by a more efficient use of the drive beam. Possibly the best option is to use part of the drive beam pulse train as it comes out of the delay loop (see Fig. 2).

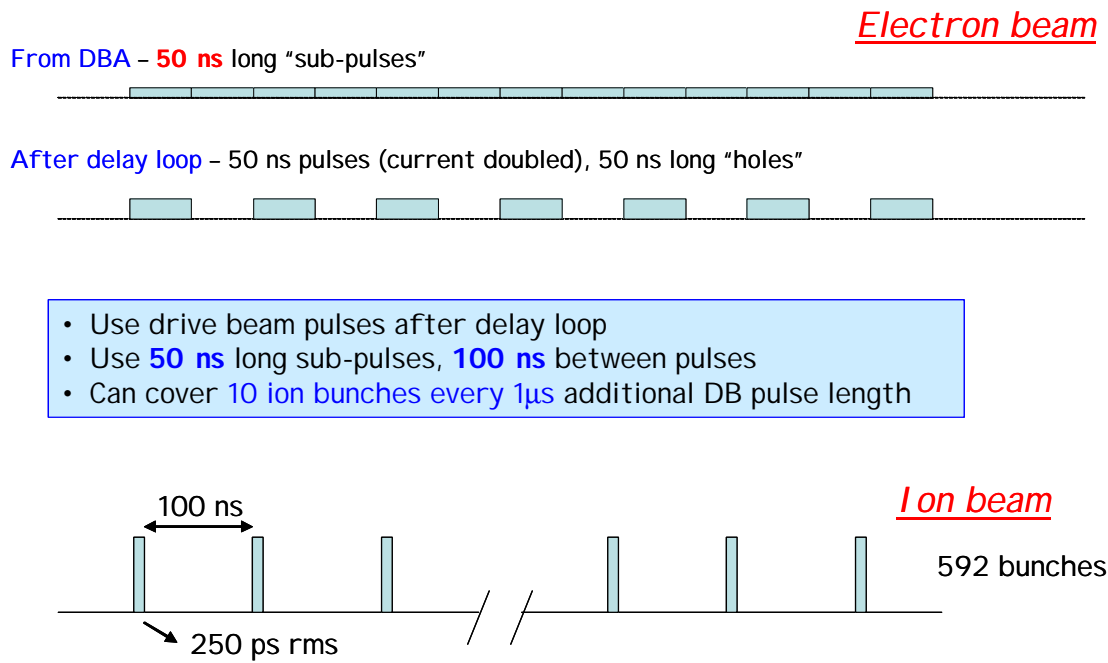

Figure 2. Matching the time structure of the drive beam and the LHC ion beam

Assuming the still tentative new CLIC parameter list found in [11], one see that at this stage a drive beam is composed by 70 ns pulses, spaced by 140 ns. Assuming for instance the use of an alternative, slightly shorter delay loop this can be brought easily to 50 ns long pulses, spaced by 100 ns. Each pulse can then be used to provide an RF pulse useful to accelerate an electron pulse. The total number of pulses available depends on how large a fraction of additional power one is willing to use. As an example, we consider using 10 µs, i.e., 10 % of the total CLIC drive beam pulse, which will provide $n_b = 100$, increasing the luminosity by the same factor.

### 3.2 Linac and Injector Parameters

The average beam current over the 50 ns drive beam electron pulses is 11 A, and the beam energy 2.4 GeV, for a peak power of 26 GW. Assuming a rather modest extraction efficiency of 75 % from the drive beam to the 15 GHz RF, a total of 20 GW peak RF power would be available for the FEL linac.

A first evaluation of the linac parameters can be based on the CTF3 SICA (slotted iris constant aperture) accelerating structures, operating at 3 GHz. Such structures have a group velocity $v_g$ = 0.0375 c. Scaling all geometrical parameters linearly with the frequency **n** from 3 GHz to 15 GHz, the group velocity does not change, and a structure length of $L$ = 0.566 m is needed to get a filling time **t** = $v_g$ $L$ = 50 ns. Properly scaling the SICA shunt impedance $r$` with frequency, one gets $r$` = 9.4 $10^3$ Ω/m. The input power per structure $P$ needed to reach a given average accelerating gradient <$E$> can be obtained using the formula:

$$\langle E \rangle = \sqrt{\mathbf{w} \frac{r'}{v_g} P \frac{1-\exp(-\mathbf{a}\, L)}{\mathbf{a}\, L}} \quad (13)$$

where **w** = 2π **n**, **a** = **w**/(2 $Q$ $v_g$). In our case $P$ = 60 MW in order to reach <$E$> = 60 MV/m. A total of about 100 structures will be needed for acceleration up to 3.4 GeV. The total active length of the linac will be 57 m, and assuming a fill factor 0.75 the overall linac length will be 75 m. The total peak power needed is 6 GW (about 80 MW/m), well below the supposed available RF power of 20 GW. The design of Power Extraction and Transfer Structures (PETS) operating at 15 GHz and adapted to the previous parameters must of course be addressed, but does not seem forbiddingly difficult. The 15 GHz accelerating structures will experience a burst of 50 ns long RF pulses, for a duration of several microseconds. Therefore, pulsed surface heating can be a potential problem, limiting a the maximum length of drive beam pulse that can be used.

For simplicity, an injector similar to TTF can be considered here. In particular, it can be based on a photocathode RF gun, providing a bunch charge of 1 nC and a normalized emittance of ~ 1 π mm mrad. The gun RF frequency can be any sub-harmonic of 15 GHz, in the 1 to 3 GHz range. Pre-acceleration to about 100 MeV at such a low frequency is probably needed, to minimize the energy spread while the bunches are still long. At this point a first bunch compression stage will be needed. One or more compression stages will then be needed to reach a final bunch length of the order of 50 μm, and will be interleaved with 15 GHz acceleration. The exact value of the final energy spread will depend on the exact arrangement of the compression stages. An rms value of ~ $0.5\ 10^{-3}$ seems anyway to be possible.

**3.3 FEL Parameters, Final Luminosity Evaluation**

The optimization of the FEL parameters requires a series of 3-D simulations of the FEL process, needing as well a detailed knowledge of the initial electron beam parameters. However, guidelines can be obtained using a simple 1-D model with a semi-analytical evaluation of 3-D effects (see, e.g., ref. [15]). The optimization criteria can be slightly different than usual, since parameters like the total number of photons and the photon beam transverse size play an important role in determining the luminosity.

We have also assumed that the FEL frequency is tuned by changing the beam energy, keeping the wiggler period and magnetic field constant. At 4 nm, we considered a beam energy of 1.1 GeV, and a wiggler field and period of 0.43 T and 2.5 cm, respectively. An average β-function of 2 m, obtained by additional focusing in the wiggler, should be not far from the optimum, yielding a saturation length below 20 m, and a total number of photons $n_g = 2.7\ 10^{13}$. At 0.4 nm, the beam energy considered is 3.5 GeV, resonant for the same wiggler period with a field, but with different additional focusing, β = 6 m. The saturation length is about 50 m and $n_g = 2.5\ 10^{12}$. The rms beam size is 30 μm in both cases. These values must be taken as indicative values, to be confirmed by 3D simulations. The values of luminosity obtained using such results, and in the case of $n_b$ =100, discussed before, is $L = 2.4\ 10^{29}$ cm$^{-2}$s$^{-1}$ at 4 nm wavelength and 10 times less at 0.4 nm. This luminosity will give very high statistics and event rates. Table 2 shows the needed FEL energies, resonant and average cross sections and event rates for CLIC*LHC based FEL-Nucleus collider.

## 4 Physics Search Potential of CLIC-LHC Based FEL-Nucleus Collider

Decay widths, spin and parity of a lot of excited levels are not determined by NRF methods. FEL-Nucleus collider will give opportunity to measure these quantities.

### 4.1 Measure of unknown decay width

FEL-nucleus colllider will give opportunity to determine unknown decay width using known ones. Indeed, decay width can be estimated using following relation:

$$\Gamma_{(1)} \cong \frac{E_g^{(1)}}{E_g^{(2)}} \cdot \frac{N_{(1)}}{N_{(2)}} \cdot \frac{\sigma_{(2)res}}{\sigma_{(1)res}} \cdot \Gamma_{(2)} \qquad (14)$$

where index 1 (2) corresponds to level with unknown (known) decay width and N is the number of observed events. For $N_{(1)}$ we can use 100 events per second as the observation limit.

As an example, if we use 4.085 MeV level of Pb ($\Gamma = 0.783$ eV, $\sigma_{res}= 7.3 \cdot 10^{-22}$ cm$^2$ and $N_{(2)}= 6.1 \cdot 10^5$ events/s), the upper limit for decay width of 11.45 MeV level ($\sigma_{res} = 0.5 \cdot 10^{-22}$ cm$^2$ for J = 1) is determined as $\Gamma = 5.24 \cdot 10^{-3}$ eV$\cdot$ ($N_1/100$).

### 4.2 Determination of the spin

In fixed target experiments the spin of excited nucleus can be determined using angular distribution of the emitted photons. In our case, this angular distribution will be transferred to the energy distribution in the laboratory frame. For spin 1 and 2 cases, angular distributions in the rest frame are given by

$$W(\theta) = \frac{3}{4}(1 + \cos^2\theta)$$

$$W(\theta) = \frac{5}{4}(1 - 3\cos^2\theta + 4\cos^4\theta) \qquad (15)$$

respectively. In the laboratory system, these distributions will be seen by detector as energy distributions (for $\gamma \gg 1$)

$$W(x) = \frac{3}{4}(x^2 - 2x + 2)$$

$$W(x) = \frac{5}{4}(4x^4 - 16x^3 + 21x^2 - 10x + 2) \qquad (16)$$

where $x = E_\gamma / \gamma_A \omega$. Here, x varies from 0 to 2 (x = 0 corresponds to $\theta = 180°$ and x = 2 corresponds to $\theta = 0°$). Figure 3 shows the x dependence of normalized energy distributions. Taking into account the high statistics provided by the proposed experiment, it is obvious that different spin values can be easily identified.

**4.3 Determination of the parity**

In the nucleus rest frame, parity $\pi$ of spin 1 dipole excitations of a nucleus with $0^+$ ground state can be determined using a linearly polarized FEL beam by measuring

$$\sum \frac{W(\theta = 90°, \varphi = 0°) - W(\theta = 90°, \varphi = 90°)}{W(\theta = 90°, \varphi = 0°) + W(\theta = 90°, \varphi = 90°)} = \pi_1 = \begin{cases} +1 \text{ for } J^\pi = 1^+ \\ -1 \text{ for } J^\pi = 1^- \end{cases} \qquad (17)$$

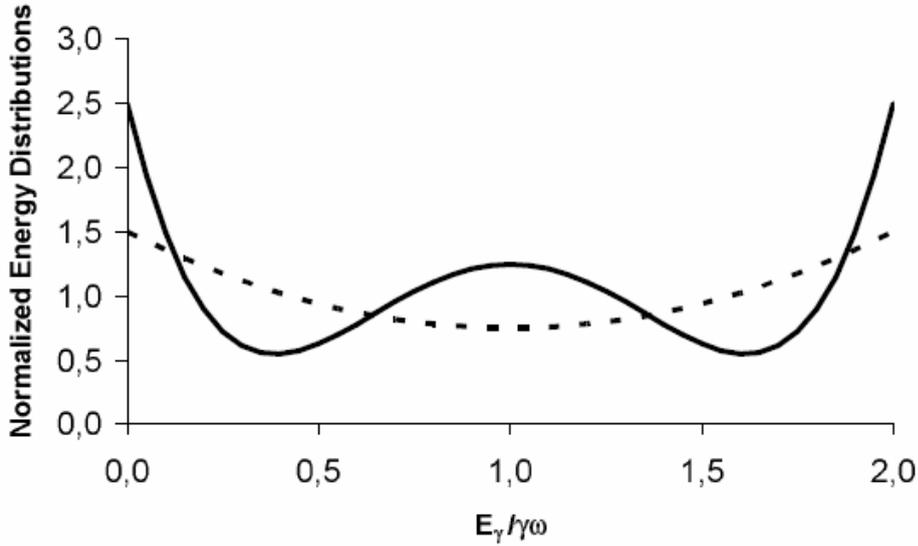

Figure 3. Normalized energy distribution of photons emitted by spin 1 (dashed curve) and spin 2 (continuous curve) excitations of the nucleus.

In our case, $\theta = 90^\circ$ corresponds to photons with $E_\gamma = \gamma_A \omega$, which are emitted at $\eta_{max} = 1/\gamma_A$, where $\eta$ is the angle between the emitted photon and initial nucleus beam direction. The azimuthal angle $\varphi$ with respect to the (horizontal) polarization plane of the $\gamma$ beam is unchanged when transferred to the laboratory frame. If the detector is placed at a distance 100 m from the interaction point, the spot size of emitted photons will have a radius about 1 m. Therefore, the measurement of parity can be made easily.

### 4.4 Pb example

Main characteristics of the $^{208}$Pb excitations with low spin (J=2), observed in $(\gamma,\gamma')$ reactions are presented in the Table 2 [16]. Let us mention that there are a lot of other (J=2) levels of the $^{208}$Pb which are observed in different reactions, such as (e,e') etc., these levels are not observed by traditional NRF methods, but most of them will be seen at FEL-nucleus collider. One can see from the Table 2 that decay width, spin and parities of a number of levels observed $(\gamma,\gamma')$ reaction cannot be determined by traditional NRF experiments.

The needed FEL energies ($w_{FEL}$), resonant and average cross-sections and event numbers per second for CLIC-LHC based FEL-Pb collider are given on the last four columns of the Table 2. One can see that due to huge events rates, FEL-nucleus collider will give opportunity to determine main characteristics of excited nuclei in a very short time period comparing with traditional NRF methods.

Table 2. Main characteristics of some of the $^{208}$Pb nucleus excitations.

| $E^*$ (MeV) | $G$ (eV) | $J^p$ | $w_{FEL}$ (keV) | $s_{res}$ (cm$^2$) | $s_{ave}$ (cm$^2$) | R/s |
|---|---|---|---|---|---|---|
| 4.0852 | 7.83·10$^{-1}$ | 2+ | 0.694 | 7.3·10$^{-22}$ | 1.39·10$^{-24}$ | 5.2·10$^5$ |
| 4.8422 | 99.72·10$^{-1}$ | 1- | 0.823 | 5.23·10$^{-22}$ | 1.07·10$^{-23}$ | 4.0·10$^6$ |
| 5.2926 | 13.16·10$^0$ | 1- | 0.899 | 2.62·10$^{-22}$ | 2.62·10$^{-24}$ | 2.4·10$^6$ |
| 5.5122 | 32.91·10$^0$ | 1- | 0.937 | 2.41·10$^{-22}$ | 0.14·10$^{-22}$ | 5.2·10$^7$ |
| 5.8461 | 11.54·10$^{-1}$ | 1+ | 0.993 | 21.5·10$^{-22}$ | 4.25·10$^{-24}$ | 1.5·10$^5$ |
| 5.9480 | 10.12·10$^{-1}$ | 1- | 1.011 | 2.07·10$^{-22}$ | 3.52·10$^{-25}$ | 1.3·10$^4$ |
| 6.2640 | 10.12·10$^{-1}$ | 1- | 1.064 | 1.87·10$^{-22}$ | 3.02·10$^{-25}$ | 1.1·10$^4$ |
| 6.3117 | 36.56·10$^{-1}$ | 1- | 1.072 | 1.06·10$^{-22}$ | 1.06·10$^{-24}$ | 3.9·10$^5$ |
| 6.3628 | 10.44·10$^{-1}$ | 1- | 1.081 | 1.81·10$^{-22}$ | 2.96·10$^{-25}$ | 1.1·10$^4$ |
| 6.7205 | 10.97·10$^0$ | 1- | 1.142 | 1.62·10$^{-22}$ | 2.64·10$^{-24}$ | 0.9·10$^5$ |
| 6.9800 | 50.64·10$^{-1}$ | - | 1.186 | - | - | - |
| 7.0635 | 28.61·10$^0$ | 1- | 1.200 | 1.47·10$^{-22}$ | 5.95·10$^{-24}$ | 2.2·10$^5$ |
| 7.0834 | 14.62·10$^0$ | 1- | 1.200 | 1.46·10$^{-22}$ | 3.62·10$^{-24}$ | 1.3·10$^5$ |
| 7.2430 | 15.67·10$^{-1}$ | - | 1.231 | - | - | - |
| 7.2780 | 15.67·10$^{-1}$ | - | 1.237 | - | - | - |
| 7.2789 | 14.00·10$^{-1}$ | 1+ | 1.237 | 1.38·10$^{-22}$ | 2.65·10$^{-25}$ | 0.9·10$^4$ |
| 7.3325 | 38.71·10$^{-0}$ | 1- | 1.246 | 1.36·10$^{-22}$ | 7.18·10$^{-24}$ | 2.7·10$^5$ |
| 7.6853 | - | - | 1.306 | - | - | - |
| 10.050 | - | - | 1.708 | - | - | - |
| 10.600 | - | - | 1.801 | - | - | - |
| 11.450 | - | - | 1.975 | - | - | - |

# 5 Results and Conclusion

It is shown that a CLIC-LHC based FEL-Nucleus collider allows to perform high precision nuclear spectroscopy. The Ion beam of LHC can be set-up for different nuclei using the existing accelerator chain from Linac 3 to LHC without major hardware changes. The proposed facility satisfies all requirements of an ideal NRF source for nuclear spectroscopy and utilizes the high boost of the LHC nucleus beams.

## Acknowledgements

This work is supported by CERN, Turkish Atomic Energy Authority and Turkish State Planning Organizaton (DPT) under the Grants No 2003K120060, 2003K1201906-5 and 2002K120250.